\documentclass[twocolumn,preprintnumbers,amsmath,amssymb]{revtex4}

\usepackage{graphicx}
\usepackage{dcolumn}
\usepackage{bm}
\usepackage[utf8]{inputenc}
\usepackage[T1]{fontenc}
\usepackage{mathptmx}
\usepackage{subfigure} 
\usepackage{soul}
\usepackage{kpfonts}
\usepackage{color}
\usepackage{hyperref}
\begin{document}


\title{\textcolor{black}{The physical origins of extreme cross-polarization extinction in confocal microscopy}}

\author{Meryem Benelajla$^{1,2}$}
\author{Elena Kammann$^{1}$}%
\author{Bernhard Urbaszek$^{2}$}%
\author{Khaled Karrai$^{1}$}%
\email{ khaled.karrai@attocube.com}
\affiliation{$^{1}$attocube systems AG, Eglfinger Weg2., 85540 Haar, Germany}
\affiliation{$^{2}$Université de Toulouse, INSA-CNRS-UPS, LPCNO, Toulouse 31077, France}


\begin{abstract}
Confocal microscopy is an essential imaging tool for biological systems, in solid-state physics and nano-photonics. Using confocal microscopes allows performing resonant fluorescence experiments, where the emitted light has the same wavelength as the excitation laser. Theses challenging experiments are carried out under linear cross-polarization conditions, rejecting laser light from the detector. In this work we uncover the physical mechanisms that are at the origin of the \textcolor{black}{yet unexplained} high polarization rejection ratio which makes these measurements possible. We show in both experiment and theory that the use of a reflecting surface (i.e. the beam-splitter and mirrors) placed between the polarizer and analyzer in combination with a confocal arrangement explains the giant cross-polarization extinction ratio of $10^8$ and beyond. We map the modal transformation of the polarized optical Gaussian beam. We find an intensity “hole” in the reflected beam under cross-polarization conditions. We interpret this as a manifestation of the Imbert-Fedorov effect, which deviates the beam depending on its polarization helicity. \textcolor{black}{This implies that this topological effect is amplified here from the usually observed nanometer to the micrometer scale due to our cross-polarization dark field methods.} We confirm these experimental findings for a large variety of commercially available mirrors and polarization components, allowing their practical implementation in many experiments. 
\end{abstract}

\maketitle


\section{\label{sec:level1}Introduction}

\textcolor{black}{In optical spectroscopy experiments it is crucial to excite an emitter with a laser very close to its transition energy. Experiments under resonant excitation are essential for accessing the intrinsic optical and spin-polarization properties of large class of emitters \cite{aharonovich2016solid,hogele2012dynamic,paillard2001spin,Karrai2004,atature2006quantum}. 
Using linear cross-polarization in a confocal setup has been successfully employed as a dark-field method to carry out resonant fluorescence experiments to suppress scattered laser light, with the added benefit of high spatial resolution \cite{vamivakas2009spin,kaldewey2018far}. Resonant fluorescence experiments allow crucial insights into light-matter coupling, such as the interaction of a single photon emitter with its environment \cite{nguyen2011ultra}, with optical cavities \cite{najer2019gated} and also  studying single defects in atomically thin materials such as WSe$_2$ \cite{kumar2016resonant}. 
Dark-field confocal techniques allow developing single photon sources with high degrees of photon indistinguishability \cite{nowak2014deterministic,muller2014,scholl2019} and longer coherence \cite{kuhlmann2013charge}. In practice dark-field laser suppression is not just a spectroscopy tool, it is also a key part of more matured quantum technology systems \cite{quandela}. \\
\indent Despite many advances based on experiments in confocal microscopes with cross-polarization laser rejection, the physical mechanisms that make these experiments possible are not well understood, hampering further progress in this field. The key figure of merit is the suppression of the excitation laser background by at least six orders of magnitude. Indeed a suppression by a factor of $10^8$  \cite{kuhlmann2013dark} up to $10^{10}$ (this work) have been measured.  But this is very surprising as mirrors and beam-splitter in such a system reduce the theoretical extinction limit to the $10^3$ to $10^4$ range. \\
\indent In this work we explain the physics behind the giant enhancement of the extinction ratio by up to seven orders of magnitude that make microscopy based on dark-field laser suppression possible. The measurements of resonant fluorescence are typically performed in an epifluorescence geometry \cite{kuhlmann2013dark}, for which laser excitation and fluorescence collection are obtained through the same focusing lens. This involves necessarily the use of a beam-splitter orienting the back-reflected light containing the fluorescence towards a detection channel. In our work we identify two key ingredients that explain the giant amplification of the cross-polarization extinction ratio : (i) a reflecting surface (i.e. the beam-splitter) placed between a polarizer and analyzer, and (ii) a confocal arrangement. We demonstrate giant extinction ratios in our experiments for different mirrors (silver, gold, dielectric, beam-splitter cubes) and polarizers (Glan-Taylor, nanoparticle thin films). We demonstrate that behind this general observation lies the intriguing physics of the Imbert-Fedorov effect \cite{fedorov1955k,imbert1972calculation}, which deviates a reflected light beam depending on its polarization helicity. We discover that a confocal arrangement not only amplifies the visibility of the Imbert-Fedorov effect dramatically, taking it from the nanometer to the micrometer scale, but also exploits conveniently the symmetry of the newly observed Imbert-Fedorov modes to insure that the cross-polarized laser beam is not coupled, explaining the near complete suppression of the laser background signal. In other words, we cannot treat the spatial (i.e. modal) and polarization properties of light separately in our dark-field confocal microscope analysis. In addition to new developments in dark-field microscopy our experiments provide powerful tools to understand spin-orbit coupling of light  \cite{liberman1992spin,bliokh2015spin, RevModPhys.91.015006}, in the broader context of topological photonics \cite{bliokh2015spin, RevModPhys.91.015006}.\\
\indent In our work we setup a robust, highly reproducible experiment and derive a convenient classical formalism to investigate these remarkable effects at the cross roads of quantum optics and topological photonics.}\\
\begin{figure*}
\includegraphics[scale=.6]{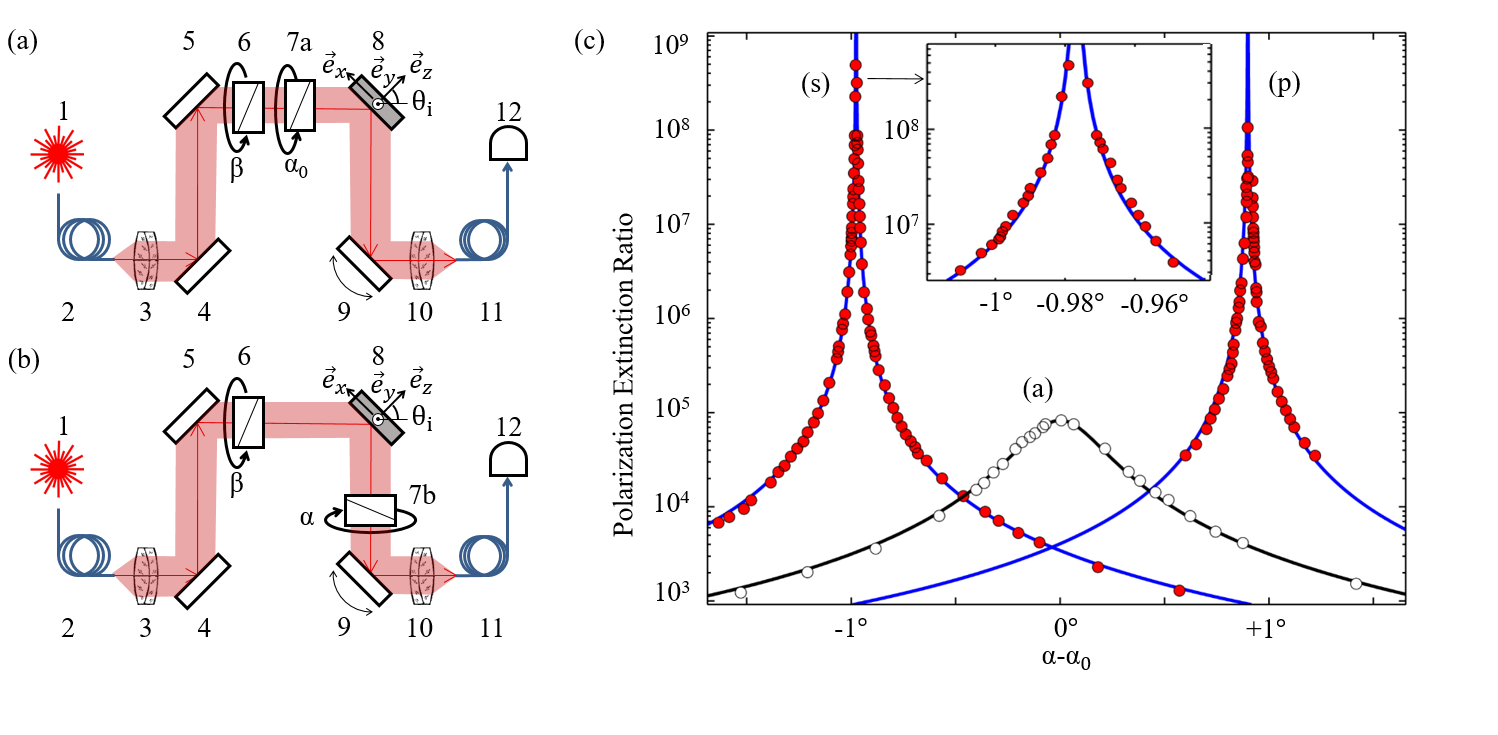}
\caption{\label{fig11} (a)(b) Cross-polarization extinction setup in a confocal microscope arrangement setup as described in the text. (c) Measured and modeled linear polarization extinction ratio for both $p$- and $s$-polarized beam around cross-polarization conditions obtained by placing the analyzer before (panel (a)) and after (panel (b)) a protected silver mirror. A giant extinction enhancement of more than 3 orders of magnitudes is obtained in configurations $s$- and $p$- with the reflecting surface placed between crossed polarizers. The inset is a magninifcation of the polarization exctinction near maximum for the $s$-polarization. An angular shift of the location of maxima of extinction is systematically found for the $s$- and $p$-polarization with respect to the reference.}
\end{figure*}
\indent The paper is structured as follows, in section \ref{setup} we introduce the experimental setup, cancellation of polarization leakage is measured and discussed in a first simplified model in section \ref{sec:level2}, the modal transformation of a reflected Gaussian beam is analyzed in section \ref{Modal}. The effect of confocal filtering is discussed in section \ref{Effect}.
\section{Confocal microscope setup}
\label{setup}
We used a simplified confocal arrangement as depicted in Fig.~\ref{fig11}a,b in order to focus on the most relevant physics leading to extreme laser rejection. A diode laser beam (1) at $\lambda= 905~nm$ wavelength is launched into a single mode fibre (2). The light emerges from the $4^{\circ}$ angled flat-polished end with a nearly perfect Gaussian beam with $ \omega_0= 2.5~ \mu m$ mode waist radius at $1/e^2$ of the maximum intensity. A diffraction limited microscope objective (3) of numerical aperture NA= 0.25 and focal length of $f=26~mm$ focused on the fiber end collimates the light into a 3 mm waist radius Gaussian beam. We choose the NA to be significantly larger than the diverging beam half-angle out of the fiber in order to preserve the Gaussian quality of the beam.\\
\indent A pair of mirrors (4,5) mounted on two axis tilt-stages allows for fine steering of the collimated beam axis. Next the beam travels to a linear polarizer (6) mounted on a piezoelectric stepping stage rotating with $20~\bm{\mu} rad$ resolution around the optical axis. The best quality commercial linear polarizers we used for this experiment showed an extinction in direct cross polarization limited to $10^5$ for nanoparticle thin film polarizer and to $10^6$ for Glan-Thomson crystal polarizers. The beam travels then towards a mirror (8), the key element of this experiment, either by passing first through an analyzing polarizer (Fig.~\ref{fig11}a) for the control measurement, or by passing through the analyzer after a reflecting surface for the test experiment (Fig.~\ref{fig11}b). We mounted the analyzer also on a piezo stepper fine rotation stage. The mirror (9) mounted on two-axis piezo controlled tilt stage steers the beam into a microscope objective (10) identical to (3) focusing the light into the core of a single mode fiber (11) identical to (2) allowing for Gaussian TEM$_{00}$ modal confocal filtering and optical detection ((12)- Si-photodiode) at the other end of the 5~m fiber cable. This confocal arrangement simulates the essential components of the resonant fluorescence confocal microscopes. The reflecting surface plane (8) at $45^{\circ}$ of incidence, defines the standard $p$ and $s$ state of polarization with projections along $\vec{e}_x$ and $\vec{e}_y$ respectively. The reflecting test surfaces in position (8) of Fig.~\ref{fig11}a,b we used in this work were commercial protected silver, aluminum and dielectric high reflectivity Bragg mirrors, evaporated gold film, as well as non-polarizing beam-splitter cubes. All such reflecting surfaces are typically used in diffraction limited confocal microscopes. The results were qualitatively very similar for all these reflecting surfaces. We choose to show here the data measured with silver mirrors only, this with the exception of data measured for comparison on a glass surface reflecting from air as discussed at the end of this publication. \\
\indent We now discuss the measurements in the configuration shown in Fig.~\ref{fig11}b, for which the reflecting test surface is sandwiched between the polarizer and the analyzer. First, the polarizer angle $\beta$ is adjusted near 0 or $\pi$/2, for $p$- or $s$-polarization respectively, while setting the analyzer angle $\alpha$ near cross-polarization at $\beta\pm \pi/2$. Then the polarizer and analyzer are subsequently finely rotated to reach maximum extinction at values $\beta$ and $\alpha$ respectively. Once the optimization reached, $\beta$ remains untouched and the analyzer in its rotator is subsequently placed before the reflecting surface just after the polarizer for our control extinction measurement (Fig.~\ref{fig11} a). The analyzer angle must be then be adjusted to a new value $\alpha_0$ in order to recover maximum nominal extinction specification inherent to the polarizers; $\alpha_0$ defines then the $p$ or $s$ reference. The extinction data measured as a function of the analyzer angle $\alpha$ in reference to $\alpha_0$ are shown in Fig.~\ref{fig11}c for the control measurement (Fig.~\ref{fig11}a) as well as for the $p$- and $s$-polarizations in the configuration (Fig.~\ref{fig11}b). Two striking observations stands out. (i) For all the tested reflectors indicated above, the extinction ratio obtained this way was enhanced beyond the $10^8$ range when the test mirror surface was sandwiched between the polarizer and the analyzer, reducing this way significantly the polarization leakage of the polarizers. (ii) The analyzer angle for maximum extinction is shifted away from $\alpha_0$ by $+ 0.898^{\circ}$ and $- 0.977^{\circ}$ for the $p$- and $s$-polarization respectively, a significant angular deviation given our resolution of about $10^{-3}$ deg. In the next section we provide a first explanation for these two striking observations.

\section{\label{sec:level2}Cancellation of polarization leakage}
Intuitively, the significant reduction of the polarization leakage field must find its root in an destructive interference effect. The first challenge towards finding an answer to our problem is to offer a model of the polarization leakage. In order to determine the light field at various planes such as after the polarizers and mirrors, we define a right hand coordinate system $\vec{p}$, $\vec{s}$ transverse to the optical beam propagation axis  $\vec{p}\times\vec{s}$ according to the definition of $p$- and $s$-polarization with respect to the plane of incidence with the test surface (8) of Fig.~\ref{fig11}b. For clarity, $\vec{s}\equiv\vec{e}_y$ is perpendicular to the incidence plane. In this section, we will test first the simplistic idea that the collimated laser beam can be approximated by a plane wave. We use a Jones matrices formalism projecting the field components along $\vec{p}$, $\vec{s}$ after each relevant optical element namely the matrix $\bar{\bar{P}}(\beta)$ of the polarizer, $\bar{\bar{M}}$ of the reflecting test surface and $\bar{\bar{A}}(\alpha)$ that of the analyzer. In this formalism an ideal linear polarizer along $\vec{p}$ and $\vec{s}$ is represented by
\begin{eqnarray}
\bar{\bar{P}}_{p_0}=\begin{bmatrix}
1 & 0 \\
0 & 0 
\end{bmatrix},
~\bar{\bar{P}}_{s_0}=\begin{bmatrix}
0 & 0 \\
0 & 1 
\end{bmatrix}~~~~~~~~~~~~~~~~~~~~~~~~~~~~~~~~~~~~~~~~~~~~~~~~
\label{eq:one}
\end{eqnarray}
We will assume now that a real physical linear polarizer along $\vec{p}$ or $\vec{s}$ represented by $\bar{\bar{P}}_{p}=\bar{\bar{L}} \bar{\bar{P}}_{p_0}$
and $\bar{\bar{P}}_{s}=\bar{\bar{L}} \bar{\bar{P}}_{s_0}$ respectively and is characterized by a polarizer leakage Jones matrix $\bar{\bar{L}}$. The assumption we are making about the physical origin of the leakage is that it is due to lossless coherent scattering such as Rayleigh scattering inclusions in the crystal. This implies that $\bar{\bar{L}}$ is unitary. The second assumption, which we verified experimentally, is that the leakage should be invariant upon an arbitrary angular rotation $\varphi$ around the optical axis $\vec{p}\times\vec{s}$, namely  $\bar{\bar{L}}= \bar{\bar{R}}(\varphi)~\bar{\bar{L}}~\bar{\bar{R}}(-\varphi)$
where the rotation matrix $\bar{\bar{R}}(\varphi)$ is given by
\begin{eqnarray}
\bar{\bar{R}}(\varphi)=\begin{bmatrix}
\cos\varphi  & -\sin\varphi \\
\sin\varphi & \cos\varphi 
\end{bmatrix}~~~~~~~~~~~~~~~~~~~~~~~~~~~~~~~~~~~~~~~~~~~~~~~~~
\label{eq:two}
\end{eqnarray}
We choose to represent the polarization leakage by a matrix
\begin{eqnarray}
\bar{\bar{L}}=\begin{bmatrix}
a  & ib \\
-ib & a 
\end{bmatrix}~~~~~~~~~~~~~~~~~~~~~~~~~~~~~~~~~~~~~~~~~~~~~~~~~~~~~~~~~~~~~~~~~~
\label{eq:three}
\end{eqnarray}
where $a^2+b^2=1$. Such a form is unitary (i.e. lossless) and invariant upon rotation. For a high quality commercially available linear polarizer $a^2\gg b^2$, which is the case in our setup since from our experiment we determine  $a^2/ b^2\cong(1.5\pm0.5)\times10^5$. This is the measured nominal leakage seen in Fig.~\ref{fig11}c. We note that the formalism can also be extended to circular polarizers, in which case $a^2\cong b^2$ and the leakage stems from the slight difference between the two terms.\\

\indent We assume an incoming laser field $\vec{E}_p$ initially $p$-polarized that we rotate at an angle $\beta$ aligning it with the polarizer such $\vec{E}(\beta)=\bar{\bar{R}}(\beta)\vec{E}_p$.
This field first traverses the leaky polarizer also rotated at $\beta$ such $\bar{\bar{P}}(\beta)=\bar{\bar{R}}(\beta)~\bar{\bar{L}}\bar{\bar{P}}_{p_0}~\bar{\bar{R}}(-\beta)$ followed by the mirror matrix $\bar{\bar{M}}$
and by the analyzer matrix rotated at an angle $\alpha$ namely $\bar{\bar{A}}(\alpha)=\bar{\bar{R}}(\alpha)~\bar{\bar{L}}\bar{\bar{A}}_{p_0}~\bar{\bar{R}}(-\alpha)$
so the field $\vec{E}$ just after the analyzer writes
\begin{eqnarray}
\vec{E}=\bar{\bar{A}}(\alpha)\bar{\bar{M}}\bar{\bar{P}}(\beta) \bar{\bar{R}}(\beta)\vec{E}_p
~~~~~~~~~~~~~~~~~~~~~~~~~~~~~~~~~~~~~~~~~~~~~~~~~~~~~
\label{eq:four}
\end{eqnarray}
The mirror Jones matrix for a plane wave writes
\begin{eqnarray}
\bar{\bar{M}}=\begin{bmatrix}
r_p  & 0 \\
0 & r_s 
\end{bmatrix}~~~~~~~~~~~~~~~~~~~~~~~~~~~~~~~~~~~~~~~~~~~~~~~~~~~~~~~~~~~~~~~~~~~~~
\label{eq:five}
\end{eqnarray}
Where $r_{p,s}$ are the complex valued Fresnel reflectivity coefficients
$r_p=(\epsilon cos\theta_i-\sqrt{\epsilon-sin^2\theta_i})/(\epsilon cos\theta_i+\sqrt{\epsilon-sin^2\theta_i})$ and $r_s=(cos\theta_i-\sqrt{\epsilon-sin^2\theta_i})/( cos\theta_i+\sqrt{\epsilon-sin^2\theta_i})$ \cite{stern1963elementary} where the test surface material enters through its complex-valued dielectric function  $\epsilon=\epsilon_1+i\epsilon_2$ or equivalently its optical constant $n^2=\epsilon$, which is tabulated for noble mirror metals \cite{johnson1972optical}.
After a lengthy but straightforward calculation, we determine the light intensity just after the analyzer 
\begin{eqnarray}
\begin{small}
\begin{aligned}
I=a^2\mid r_p \cos\alpha\cos\beta+r_s\sin\alpha \sin\beta\mid^2~~~~~~~~~~~~~~~~~~~~~~~~~~~~~~~~~~~\\
 b^2\mid r_p\cos\alpha\sin\beta-r_s\sin\alpha\cos\beta \mid^2-2ab~Im(r_pr_s^*)\cos\alpha\sin\alpha 
 \\&&\label{eq:six}
 \end{aligned}
 \end{small}
\end{eqnarray}
The polarization extinction ratio is then simply given by 1/$I$. A practical check for $p(s)$ polarized light, namely for $\beta=0(\pi/2)$ and the corresponding cross-polarization $\alpha=\pi/2(0)$ leads to the expected finite polarization leakage $I=b^2\mid r_{p/s}\mid^2$. For a hypothetical perfect mirror, $r_p=1$ and $r_s=-1$ making it in this idealized case $I=b^2$. Because the reflecting surface has real and imaginary components for $r_s$ and $r_p$, equation (6) shows that for a ''sufficiently small value'' of $b^2$ we can always find a choice of angles $\alpha,\beta$ that leads to $I=0$, cancelling this way the undesired leakage. This is always true under condition of total internal reflection which is the case of a metallic mirror in the visible and infrared range and for a typical cube beam-splitter. A ''sufficiently small value'' of depolarization to obtain perfect cancellation means in the context of our work typically  $b^2< 6\times10^{-3}$ when using a silver mirror as we will derive later in the text. The reflecting test surface in combination with the polarizers rotation act to interfere destructively with the residual rotation invariant lossless polarization leakage inherent to even best commercial linear polarizers. Conversely, for a purely dielectric surface such a glass (i.e. BK7) reflecting from the air-side, for which $r_p$ and $r_s$ are both real, no full polarization leakage cancellation was possible.\\
\indent To get a better feel for the relevant parameters at work in cancelling almost perfectly the polarization leakage we use the form $r_p=\rho_p~exp(i\varphi_p)$ and $r_s=\rho_s~exp(i\varphi_s)$. Consider high reflectivity mirrors for which $\rho_p\approx\rho_s\approx1$. Solving equation (4) for field cancellation leads to the first order in $\mid b\mid \ll 1$ to a set of two equations $\cos(\alpha+\beta)=b/\tan\Delta$ and $\cos(\alpha-\beta)=b~\tan\Delta$ where $\Delta=(\varphi_p-\varphi_s)/2$. This way both $\alpha$ and $\beta$ can be analytically calculated. In the particular case of dielectrics reflecting from the air side for which $\Delta=\pi/2$ we see already that there are no solutions. Instead we need the condition $\Delta\ne\pi/2$ which is always verified in condition of total internal reflection. For pure silver at $\lambda=905~nm$, $\varphi_p-\varphi_s=192.52^\circ$ implying $\tan\Delta=-9.12$ which in turn shows that for this particular case a solution $I=0$ exist for polarizers with leakage levels $b^2<0.012$. One more step is required to make use of these equations towards interpreting our results because we did not find any easy way to measure independently an absolute value $\alpha$ and $\beta$ to the precision required for our measurements. As explained in the previous section the value we can measure experimentally with the required accuracy is the shift $\alpha-\alpha_0$. In the reference measurement with the analyzer placed directly after the polarizer we assume that the cross-polarization condition $\alpha_0-\beta=\pi/2$ holds. 
For the test experiment, the equation $\cos(\alpha-\beta)=b~\tan\Delta$ is developed in the limit of small leakage $\mid b~\tan\Delta \mid\ll1$, so that we get $\alpha-\beta=\pi/2 \mp b~\tan\Delta$ for the near $p$ and $s$ conditions respectively. This shows that the correction to the analyzer is simply $\alpha-\alpha_0=\mp b~\tan\Delta$ corresponding to $\alpha-\alpha_0=+0.898^\circ$ with a measured leakage $1/b^2=8.3\times10^4$ and $-0.977^\circ$  with $1/b^2=9.6\times10^4$ for the $p$- and $s$- state respectively in the case of the measurement of figure 1b, for a protected silver mirror. Hence, we determine $\Delta=102.48^\circ$ and $\Delta=100.7^\circ$ for $p$ and $s$ beam respectively. Given the measured leakage limiting the nominal extinction at $1/b^2 = (1.5\pm0.5)10^5$ we determine $b = (2.7\pm0.5)10^{-3}$ which in turn allows determining $\Delta=99.7^\circ\pm1.7^\circ$ a value to be compared to the value for pure silver of $\Delta =96.27^\circ$ \cite{johnson1972optical}. 
The difference could be possibly related to the effect of the protective layer or on the purity of the silver mirror we used. This simple novel method shows that we can conveniently measure the phase shift $\varphi_p-\varphi_s$ between $p$- and $s$-polarization after reflection for metals. Our measurements performed using a high reflectivity Bragg mirror showed larger shifts $\alpha-\alpha_0=\mp 2.2^\circ$ leading to a phase shift $\varphi_p-\varphi_s=189.2^\circ$ between the $p$- and $s$-reflected components.\\
\indent The full measurements shown in Fig.~\ref{fig11}c are fitted using equation 6 accounting convincingly for the cross-polarization extinction amplification and the slight polarizer and analyzer rotation shift required to reach it. In the limit $\rho_s \approx \rho_p $, we calculate that the polarization leakage should be sufficiently small to allow perfect extinction when the condition $b^2<(1-\sqrt{\tan ^2\Delta-1}/\sqrt{\tan ^2\Delta+1})/2$ is verified. For our experimental case, this corresponds to $b^2<1.26~10^{-2}$ and for a pure silver to $b^2<6.00~10^{-3}$. Such values are in fact relatively large and thus allow realistically achieving polarization leakage cancellation for most standard commercial polarizers. \\
\indent A last practical aspect to address is the wavelength dependency of this effect. For a highly reflecting mirror, the wavelength dependency is to be found in the phase difference $\Delta (\lambda)$. As a result, the correction to the analyzer angle $\alpha(\lambda)-\alpha_0=\mp b \tan \Delta(\lambda)$ calculated for reaching maximum extinction is also a function of wavelength. Hence, we see that the polarizer angle $\alpha$ of maximum extinction shifts as a function of wavelength as $\partial \alpha/\partial \lambda=\mp b(\partial \Delta/\partial \lambda)/\cos^2 \Delta$ which can easily be evaluated using the formulae of Fresnel coefficient and the corresponding dielectric constant of the mirror relevant material. For a perfect silver mirror and a polarization leakage of $10^5$ we evaluate a chromaticity rate of  $\partial \alpha/\partial \lambda=0.0019^\circ/nm$ for a wavelength around $\lambda =905~nm$. In this particular example, keeping the analyzer angle at value of maximum extinction for $905~nm$, the wavelength could be shift by up to $\pm 10~nm$ and still keep the extinction up to a level $> 10^7$. \\

\indent At this point we could conclude the paper here as we were able to explain convincingly all the features of the enhanced polarization extinction. Our analysis has however occulted so far a crucial point, namely the experimental fact that the leakage cancellation was only measurable in a confocal arrangement, a point that is elucidated in the next chapters. More specifically, the analysis we conducted leading to the main result in equation (6) so far was done purely for a plane wave for which the Jones matrix formalism is valid. In reality however, the finite size of the collimated Gaussian laser beam imposes a finite angular wave distribution around the angle of incidence  on the mirror \cite{novotny2012principles}. The Fresnel coefficient $r_p$ and $r_s$ becomes then a function of the angular distribution \cite{novotny2012principles}. This as we will see, leads to significant geometrical depolarization effects in form of new optical modes limiting the total extinction to the $10^{4}$ range. We will see that a confocal arrangement filters away the depolarization modes and that the result of this chapter turns out to be fortuitously usable.
\begin{figure*}
\includegraphics[scale=.5]{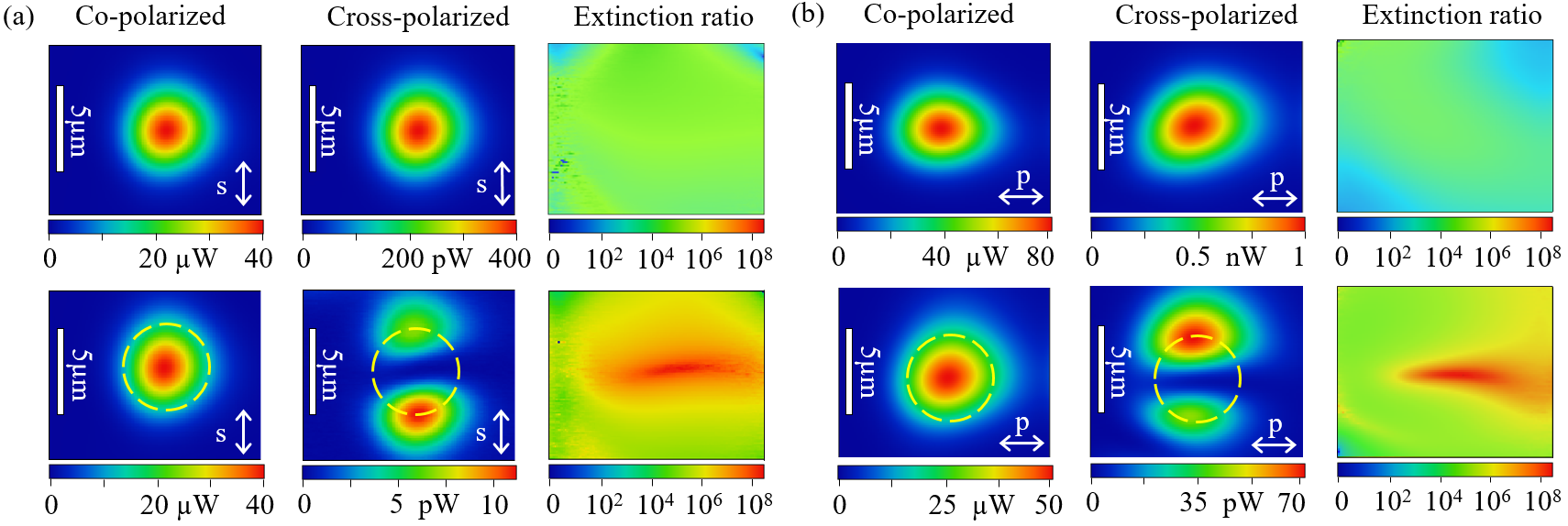}
\caption{\label{fig22} Confocal mapping of $s$- (a) and $p$- (b) laser beam in co- and cross-polarization using a scanning mirror (9). The upper and lower rows of the figures are measurements with the analyzer placed before ( Fig.~\ref{fig11}a) and after ( Fig.~\ref{fig11}b) the test reflecting surface respectively, in $s$-polarization and $p$-polarization (panel a and b). We plot the extinction ratio map by dividing, pixel per pixel, the co-with the cross-polarized data. In cross-polarization, a modal splitting along $\vec{e}_y$ above and below the plane of incidence is observed both for the $s$- and $p$-polarization. The dotted line circle is the 1/$e$ intensity level of the Gaussian distribution resulting from the confocal convolution between the focused spot and the collecting fiber mode. The location of the maxima of the modal splitting lie exactly on that circle. The diameter of this circle gives also the non-convoluted focal spot waist diameter at 1/$e^2$ focused on the collecting fiber end. The vertical white bar represents 5$\bm{\mu}$m.}
\end{figure*}

\section{Modal transformation of a reflected polarized Gaussian beam}
\label{Modal}

\begin{figure}
\includegraphics[scale=.4]{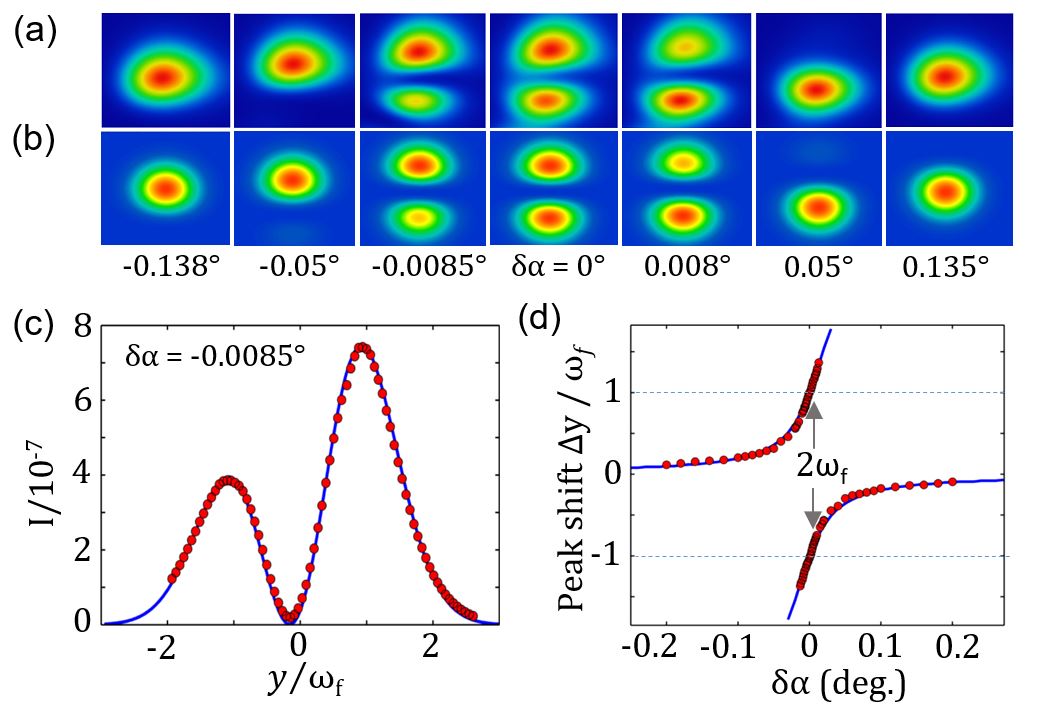}
 \caption{ \label{fig33}$p$-polarized beam reflected off a silver mirror. Measured (a) and simulated (b) evolution of the modal confocal imaging mapping through maximum extinction (c) for different analyzer angles $\delta\alpha$ as explained in text.  In (d), the beam peak-shift and splitting positions are shown in units of beam waist $\omega_f$ at focus which we modeled for our silver mirror $\Delta=102.48^\circ$and a leakage of 1/$b^2=8.3 \times 10^4$}
\end{figure}

\begin{figure}
    \includegraphics[scale=.4]{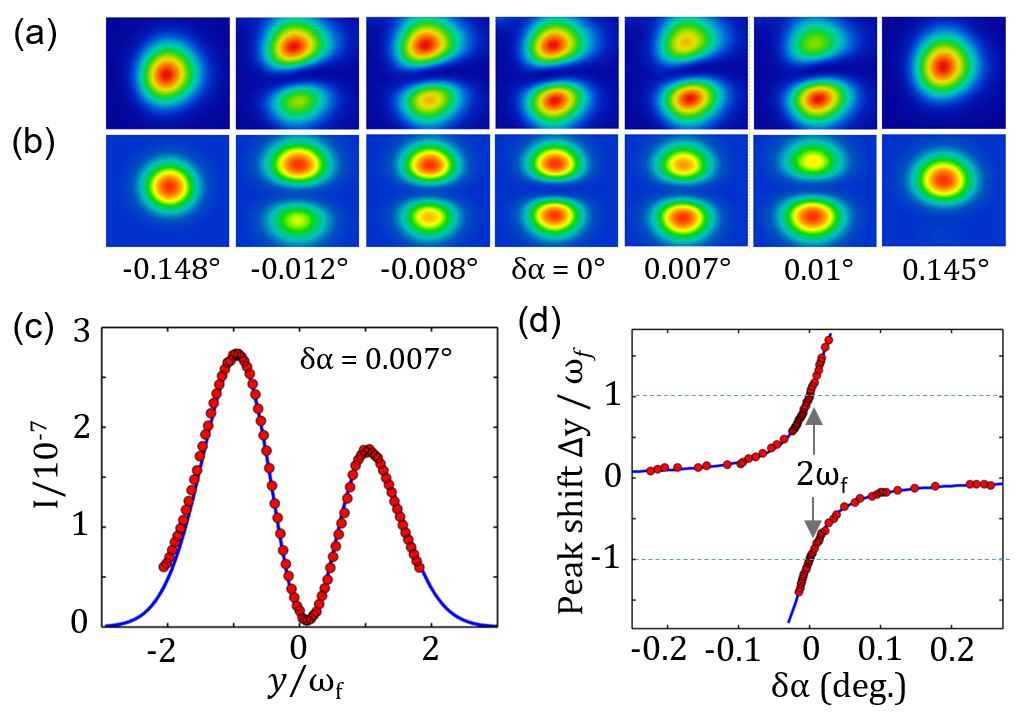}
  \caption{\label{fig44} $s$-polarized beam reflected off a silver mirror. Measured (a) and simulated (b) evolution of the modal confocal imaging mapping through maximum extinction (c) for different analyzer angles $\delta\alpha$ as explained in text. In (d), the beam peak-shift and splitting positions are shown in units of beam waist $\omega_f$ at focus which we modeled for a silver mirror $\Delta=100.7^\circ$and a leakage of 1/$b^2=9.6\times 10^4$}
\end{figure}

To find the origin of the unexpectedly high polarization rejection ratio $>10^8$, we mapped the detected intensity by scanning the spatial position of the collecting fiber in the focal plane of the focusing objective.\\
\indent In the absence of a reflecting surface between the analyzer and polarizer in cross-polarization, namely the reference configuration in Fig.~\ref{fig11}a, 
the measurements in Fig.~\ref{fig22}a,b (upper row) show a pure TEM$_{00}$  Gaussian mode
field attenuated by $8.3\times10^4$ and $9.6\times10^4$ for $p$ and $s$ polarized beam, respectively.
This level is expected for the polarizer leakage specifications. In contrast, when we place the analyzer after the reflecting test surface, the measurements show that the mode splits into two lobes distributed along $\vec{e}_y$ and located above and below the reflectivity plane. In this cross-polarized configuration, we find an intensity “hole” at the location of the optical fiber center. There the intensity extinction is slightly higher than $10^8$, a factor 100 away from our actual setup sensing-limit. We occasionally reached $10^{10}$ records in which the dark noise of the detector was in fact the limiting factor. We have experimentally verified the stability of this effect over tens of hours for $p$- and $s$- linearly polarized incident light. We observed qualitatively the same effects for incidence angles of $\theta_i$ at $9^\circ$, $22^\circ$, $25^\circ$, $30^\circ$  and $68^\circ$. We observed qualitatively the same behavior for different type of polarizers such as crystal polarizer (Glan-Taylor) and nanoparticle thin film linear polarizers, for different mirrors such as silver, gold, aluminum, dielectrics Bragg reflectors and non-polarizing beam-splitter cubes, attesting to the robustness of this effect.\\
\indent To get a feel for the measured modal transformation for $p$- and $s$-, we measured and showed in Fig.~\ref{fig33} and Fig.~\ref{fig44} the evolution of the confocal light intensity maps for different analyzer rotation angles variation $\delta\alpha$ around the symmetrically split mode.
Fig.~\ref{fig33}d and \ref{fig44}d show quantitatively for $p$- and $s$- polarizations the measured positions of beam-peak shifts along $\vec{e}_y$ and splitting above and below the plane of incidence as a function of $\delta\alpha$. We observed a very similar behavior for beam-splitter cubes typically used in the resonant fluorescence setup such as in reference \cite{vamivakas2009spin, kaldewey2018far, kuhlmann2013dark,kuhlmann2013charge}, with the difference however that equivalent figure looks instead mirrored with respect to the axis $\Delta y=0$. In all cases,  such split-lobes intensity distribution is very reminiscent of a TEM$_{01}$ Hermit-Gaussian mode. \\
Figure \ref{fig33}.c and  \ref{fig44}.c  show on careful inspection that the minima of intensity or maxima of extinction do not occur exactly at $y=0$ but instead are very slightly displaced symmetrically along y for both the $p$- and $s$- polarization. This mean that there is not a single position of the fibre location $y$ that can lead to a maximum extinction for both $p$- and $s$- polarization at the same time. This is also clearly seen in Fig.~\ref{fig11}.c for which the extinction is beyond the $10^9$ range for the $s$- polarization and $10^8$ for $p$- in that particular measurement. This observation suggest clearly that the novel modes along $y$ do seem to assist in boosting the extinction well beyond the $10^8$ level. We have reached a record level of $10^{10}$ for which the limiting factor was the dark noise of our detector. The challenge in such experiment is to have a polarization rotator that enable stepping with small enough rotation angles.  
At this point we need to find out why (i) the confocal arrangement enables the dramatic extinction enhancement as seen in Fig.~\ref{fig11}.c, and (ii) why does the beam shift and split at cross-polarization in  Fig.~\ref{fig22} (lower row) and this always above and below the plane of incidence.\\
\indent To answer these questions we need to first model the spatial field distribution $\vec{\mathcal{E}}_{fx,y}$ at the focal plane of the focusing lens just before the collecting single mode optical fiber and then use the collecting fiber as the confocal Gaussian filter function porting the light to the detector. The finite size beam before the mirror results from a Gaussian-weighted superposition of plane-waves propagating along an angular distribution $\roarrow{k}/k_0= u\roarrow{p}+v\roarrow{s}+w\roarrow{k_0}/k_0$ very narrowly centered around $\vec{k_0}$ the wave vector along the optical axis with $k_0=2\pi/\lambda$. In the paraxial approximation $u$ and $v$ are both $\ll1$, so they represent the angular spread of the collimated beam. The focusing lens transforms each plane-wave field $\vec{E}_{u,v}$ of the angular distribution  into a field density $\vec{\mathcal{E}}_{fx,y}$ in the focal plane, hence the beam reaching the focal plane at distance $f$ result from a coherent superposition of all such focused components. Assuming that all these waves are paraxial and applying the Fresnel approximation \cite{goodman1968introduction} one can establish that 
\begin{eqnarray}
\vec{\mathcal{E}}_{fx,y}=\frac{-i}{\lambda f} \exp+ ik_0f\iint \limits_{-\infty}^{+\infty}\vec{E}_{u,v}~exp+ik_0(xu+yv)\,du\,dv~
\label{eq:seven}
\end{eqnarray}
The integrals run normally within the maximum boundaries -1 and 1 for $u$, $v$, but for mathematical convenience they are extended to infinities as this does not affect the result in a paraxial approximation, namely because $\vec{E}_{u,v}$ vanishes rapidly when $u$, $v$ are no longer much less than unity. In what follows we will drop the propagation phase term $\exp+ ik_0f$ as we from now on just concern with establishing the field at focal plane only. The next step is to obtain the angular distribution ${\roarrow{E}_{u,v}}$. Such problem was modeled for a Gaussian field distribution in \cite{aiello2007the} by  Aiello and Woerdman. We derive here a simplified version conveniently describing the essential physics needed to model our observations. We begin with the field just before the polarizer for which we assume a linearly polarized Gaussian-field normalized angular distribution $\roarrow{E}_{0uv}$
\begin{eqnarray}
\roarrow{E}_{0uv}=\frac{E_0}{\pi\theta_0^2}~exp-\frac{u^2+v^2}{\theta_0^2}\begin{bmatrix} \cos\beta  \\ \sin\beta \end{bmatrix}~~~~~~~~~~~~~~~~~~~~~~~~~~~~~~~~~~
\label{eq:eight}
\end{eqnarray}
The mode divergence $\theta_0=2/(k_0\omega_0)=\omega_0/l$ results from the finite size of the collimated laser beam with beam radius $\omega_0\equiv3mm$, and the Rayleigh range $l=k_0\omega_0^2/2$ is $10^4$ m, a value much larger than the size of our experimental setup allowing us to ignore the role of beam propagation up to the focusing lens. With this convention a $p(s)$ polarized light is obtained at $\beta=0(\pi/2)$.\\
 When the beam reflects off the test surface, each plane-wave component acquires an angle dependent Fresnel reflection coefficient $r_{p,uv}$ and $r_{s,uv}$ that are  function not only of $\theta_i$ but also of $u$, $v$ \cite{novotny2012principles}.\\
Consequently, for each plane-wave component, we choose a coordinate system $\vec{e}_p$, $\vec{e}_s$, $\vec{k}/k_0$ that defines a local incidence plane for that wave. The longitudinal basis vector is $ \vec{k}/k_0$ and the transverse ones are $\vec{e}_s=  \vec{k}/k_0 \times \vec{e}_z $ and $\vec{e}_p=( \vec{k}/k_0 \times \vec{e}_z )\times (\vec{k}/k_0)$ in the $s$- and $p$- planes respectively. To obtain the reflectivity of the mirror for each plane wave, we determine first the weights of $p$- and $s$- field-components, given by the weighted projections $r_{p,uv}(\vec{e}_p\cdot\vec{E}_{0uv})$ and
$r_{s,uv}(\vec{e}_s\cdot\vec{E}_{0uv})$. We determine then the resulting reflected field transverse field along the corresponding reflected basis $\vec{e}_{s,R}= \vec{k}_R/k_0 \times \vec{e}_z$ and $\vec{e}_{p,R}=( \vec{k}_R/k_0\times \vec{e}_z)\times (\vec{k}_R/k_0)$ such $\vec{E}_{uv}=r_{p,uv}(\vec{e}_p \cdot \vec{E}_{0uv})\vec{e}_{p,R}+r_{s,uv}(\vec{e}_s\cdot\vec{E}_{0uv})\vec{e}_{s,R}$. Here $\vec{k}_R$ is the mirrored wave vector after reflection.
In the paraxial limit, for a beam impinging, the Fresnel coefficients are developed to the first order in $u$ around $\theta_i$ and $v$ around 0, giving $r_{p,uv}=r_{p}+u~\partial r_p/\partial \theta_i$ and $r_{s,uv}=r_{s}+u~\partial r_s/\partial \theta_i$. The first order derivatives $\partial r_{p/s~uv}/\partial v $ in the $s$-plane vanish both for $r_p$ and $r_s$ leaving just derivative
$r_p^{'}=\partial r_p/\partial \theta_i$ and $r_s^{'}=\partial r_s/\partial \theta_i$. We calculate the components of the incoming and reflected basis vectors $\roarrow{e}_p$, $\roarrow{e}_s$, $\roarrow{k}/k_0$ and $\roarrow{e}_{p,R}$, $\roarrow{e}_{s,R}$, $\roarrow{k}_R/k_0$ in the paraxial limit $u, v << \theta_i$ . After a lenghty but straightforward calculation we obtain the reflected field distribution after the mirror for each angle $u, v$. We express the result conveniently in terms of Matrix notation $\vec{E}_{uv}=\bar{\bar{M}}_{u,v}~\vec{E}_{0uv}$ where
\begin{eqnarray}
\bar{\bar{M}}_{u,v}=\begin{bmatrix}
r_p  & 0 \\
0 & r_s 
\end{bmatrix}+u\begin{bmatrix}
r'_p  & 0 \\
0 & r'_s 
\end{bmatrix}+v\frac{r_p+r_s}{\tan\theta_i}\begin{bmatrix}
0 & -1 \\
1 & 0 
\end{bmatrix}~~~~~~~~~~~~~~~~~~~
\label{eq:nine}
\end{eqnarray}
Upon inspection of the expression (9) for symmetries we see now that the reflectivity Jones matrix transforms an impinging perfect Gaussian mode, such as equation (8), into the sum of TEM$_{00}$, TEM$_{01}$ and TEM$_{10}$ Hermit-Gauss modes. The indices for TEM$_{nm}$ indicate the number of nodes along the $\vec{p}$ and $\vec{s}$ direction repectively. The first term in the right hand side is the normal test surface reflectivity we used in the first part of this paper. The second term is responsible for generating a TEM$_{10}$ mode along $\vec{p}$ in the plane of incidence. This term is in fact responsible for the Goss-Hänchen effect \cite{goos1947neuer}, as it has its physical origin in the angular dispersion of the reflectivity terms at $\theta_i$. Here the different plane-wave components acquire slightly different phases upon reflectivity shifting the beam in the plane of incidence. Because this matrix is diagonal, we see that for a perfect $p$- or $s$-polarization the Goos-Hänchen effect does not contribute to depolarization.The third term, the most relevant to this work, is responsible for generating an out of plane-of-incidence TEM$_{01}$ mode with two lobes along $\vec{e}_y$. 
This term is the physics responsible for the Imbert-Fedorov effect \cite{fedorov1955k,imbert1972calculation, onoda2004, bliokh2015spin} known to deviate a reflected light beam above or below the plane of incidence depending on its right handed or left handed polarization helicity. The calculation detailed above shows that this term originates purely from geometrical projections in which the gradual phase-shift gained by each plane-wave component upon reflection, sums to a cross-diagonal matrix that mixes the $p$- and $s$- phase shifted reflected plane wave components. Consequently this term is responsible for an intrinsic reflectivity induced depolarization for $p$- and $s$-polarization even when using ideally perfect polarizers. Because of the purely geometrical projections nature of the argumentation, compelling connections between the Imbert-Fedorov effect, Berry’s phase and spin-hall effect of light are discussed in the literature \cite{onoda2004, bliokh2015spin}. Because of the direct proportional dependency of this matrix on the angle $v$ and in particular its sign, it creates a TEM$_{10}$ mode asymmetric along $\vec{e}_y$, adding/suppressing field to/from the symmetric main mode displacing this way its weight above or below the plane of incidence depending on its helicity. This can be easily verified using a circular polarization version of equation (8) with the Jones matrix equation (9). From this simple derivation, it is worth appreciating that in the paraxial approximation equation (9) express both  Goos-H{\"a}nchen and Imbert-Fedorov effects in an elegant and compact way. At this point, we can see from a symmetry argument that our confocal arrangement enhances cross-polarization extinction. Without the confocal arrangement, the extinction would have been naturally limited in the $10^4$ range in our experiment as we will discuss in the next chapter.
\\
\indent In the following step we express the field distribution transmitted through the polarizer, the mirror and analyzer at the back aperture of the focusing lens $\vec{E}_{uv}=\bar{\bar{A}}(\alpha) \bar{\bar{M}}_{u,v} \bar{\bar{P}}(\beta)\vec{E}_{0uv}$
used in the Fourier transform equation (7). Before providing the general solution, we get first a feel for the physical parameters governing the Imbert-Fedorov cross-polarized mode. For this, we consider the special case of a $p$- or $s$-polarized light impinging on the mirror and subsequently analyzed in cross-polarization configuration. Here, only the third matrix on the r.h.s of equation (9) is relevant, all other terms cancel. 
After some algebra we obtain the cross-polarized field for the $p$($s$) incident light
\begin{eqnarray}
\vec{\mathcal{E}}_{f\perp}=\frac{\pm E_0}{\pi \omega_{f}^2} \frac{r_p+r_s}{\tan\theta_i}\frac{y}{l_f}~exp-\frac{x^2+y^2}{\omega_{f}^2}\vec{e}_s~~~~~~~~~~~~~~~~~~~~~~~~~~~~~
\label{eq:ten}
\end{eqnarray}
here the focused spot waist radius $\omega_{f}=\lambda f/\pi\omega_0 \simeq 2.5\mu m$ is the fiber Gaussian mode size and $l_f=k_0 \omega_f^2/2$ corresponding Rayleigh range, in our case $21.7\mu m$. We notice that the field is an antisymmetric function of $y$ with two lobes with opposite phase located at $y=\pm \omega_f/\sqrt{2}$ above and below the plane of incidence. The field peak intensity normalized to the maximum of co-polarized peak intensity is proportional to the ratio $\omega_f/l_f$ which is nothing else than $\theta_f=2/(k_0\omega_f)$ the half cone angle of the focused beam which in our experiment is the numerical aperture of the single mode fiber. An important result emerges, namely that the lens amplifies dramatically the Imbert-Fedorov mode field strength. It is easy to show that the amplification factor of the light intensity is $\omega_0^2/\omega_f^2$ when comparing the peak strength just after the reflecting test surface (i.e $\omega_0=3mm$) and at the focal plane. In our case we obtain an amplification of $3mm/2.5\mu m =1200$. This is the reason for which we can detect this mode so clearly in a confocal configuration. \\
\indent The other essential physical parameter that governs the Imbert-Fedorov mode field intensity is the sum $r_p+r_s$ given by the material reflecting properties. To get a more physical insight we use the representation $r_p=\rho_p~exp(i\varphi_p)$ and $r_s=\rho_s~exp(i\varphi_s)$ that can be conveniently symmetrized using $\rho_s=\rho+\delta\rho/2$, $\rho_p=\rho-\delta\rho/2$ for the reflectivity and $\Delta=(\varphi_p-\varphi_s)/2$ the phase difference. This way we obtain
\begin{eqnarray}
r_p+r_s=2\rho \cos\Delta+i\delta\rho \sin \Delta~~~~~~~~~~~~~~~~~~~~~~~~~~~~~~~~~~~~~
\label{eq:eleven}
\end{eqnarray}
to within a constant proportional phase term $exp i(\varphi_p+\varphi_s)/2$ identical for all modes of equation 9. We now see that the difference $\varphi_p-\varphi_s$ governs the intensity and the phase of the Imbert-Fedorov mode. For instance in the case of the air side reflectivity off a perfect dielectric we have $\varphi_p-\varphi_s=\pi$, hence $r_p+r_s=i\delta\rho$, so the Imbert-Fedorov mode field intensity is directly proportional to the pure differential reflectivity between the $p$ and $s$ waves. In contrast, for dielectrics under total internal reflectivity and for metals we have $\delta\rho\approx0$ and $\rho\approx1$ so that $r_p+r_s=2 \cos\Delta$. In this case the strength of the depolarizing mode is fully governed by the phase difference $\varphi_p-\varphi_s$. We conclude that mapping the Imbert Fedorov mode fields in a confocal microcopy setup provides a direct and sensitive access to the differential reflectivity amplitude and phases of a reflecting surface.
To move forward with our analysis on the more general case we have performed the Fourier optics transformation equation (7). The result is that the field image at focal plane for the test experiment with the mirror placed between the polarizers is
$\vec{\mathcal{E}}_{fxy}=\bar{\bar{A}}(\alpha)\bar{\bar{M}}_{x,y}\bar{\bar{P}}(\beta)\vec{\mathcal{E}}_{0x,y}$
where the effective reflectivity Jones matrix is given by
\begin{eqnarray}
\bar{\bar{M}}_{x,y}=\begin{bmatrix}
r_p  & 0 \\
0 & r_s 
\end{bmatrix}+i\frac{x}{l_f}\begin{bmatrix}
r'_p  & 0 \\
0 & r'_s 
\end{bmatrix}+i\frac{y}{l_f}\frac{r_p+r_s}{\tan\theta_i}\begin{bmatrix}
0 & -1 \\
1 & 0 
\end{bmatrix}~~~~~~~~~~
\label{eq:nine}
\end{eqnarray}
The spatial distribution of the field $\vec{\mathcal{E}}_{0x,y}$ results from the lens transforming the unperturbed linearly polarized laser field angular distribution of equation (8) into a spatial normalized distribution now at the focal point such
\begin{eqnarray}
\vec{\mathcal{E}}_{0x,y}=\frac{-iE_0}{\pi\omega_f^2}~exp-\frac{x^2+y^2}{\omega_f^2}\begin{bmatrix} \cos\beta  \\ \sin\beta \end{bmatrix}~~~~~~~~~~~~~~~~~~~~~~~~~~~~~~~~~~
\label{eq:eight}
\end{eqnarray}
This result turns out to be within a Gouy phase $-i$ at focus, the one discovered in a different context in the insightful and pioneering work of Aiello et. al. \cite{aiello2008role}. In their work the authors provided within a paraxial approximation, a complete analytical solution for the field distribution of a single mode Gaussian beam reflected off a mirror. The essential finding from our work is that the confocal arrangement transforms the collimated beam waist $\omega_0$ and Rayleigh length $l$ and of Aiello et. al. \cite{aiello2008role} field distribution at the mirror plane, into $l_f$ and $\omega_f$ in our case. This result appears benign at first but as discussed earlier amounts to a sizeable amplification of the weak mode intensity in proportion to $\omega_0^2/\omega_f^2$. Aiello et. al. \cite{aiello2008role} showed that it is the finite size of the beam at the reflecting surface that are responsible for the additional field terms that affect the initial Gaussian mode. From our work it is becoming now clear here that using a confocal arrangement, the size of the beam at the mirror is not relevant to our measurements but rather the size of the focused beam that plays a crucial role. This fact, at first non-intuitive, provides a valuable advantage to explore experimentally the cross-polarization geometry with sufficient sensitivity and a very fine spatial resolution. In particular, high extinction cross-polarization extinction is kin to the “weak measurement procedure” of Aharonov et.al. \cite{aharonov1988result,duck1989sense} which we extend here to a confocal arrangement enabling the added benefit of spatial resolution. Recent literature \cite{bliokh2015spin} provides an interpretation for the depolarization as resulting from an effective spin-orbit interaction of light occurring at the mirror surface manifesting itself in the form of a spin hall effect of light \cite{hosten2008observation}. In the work, we restrict ourselves to a purely modal interpretation and leave the discussion concerning spin-orbit aside.

\section{Effect of confocal spatial filtering}
\label{Effect}

The one final point we need to address to get a full quantitative interpretation of our experiment is to address the effect of the confocal filter function of the collecting fibre. The single mode fiber collects and ports the field to the photo-detector, it does this however by acting as a Gaussian spatial filter. For our symmetric setup shown in Fig.~\ref{fig11} we illuminate and collect light with a single mode fibers of identical mode size, and with identical collimating and focusing lenses. The spatial filtering is a convolution between the field spatial distribution at focal plane and the fiber Gaussian mode amounting to a detected field strength $\vec{{E}}_D$ location $x_0$, $y_0$ with respect to the optical axis. The results of the calculation for the field are the following. First we get the mapping of the reference field without use of polarizer and reflecting surface as seen by the detector
\begin{eqnarray}
\vec{{E}}_{D0,0}=-\frac{iE_0}{2}~exp-\frac{x_0^2+y_0^2}{2\omega_f^2}\begin{bmatrix} \cos\beta  \\ \sin\beta \end{bmatrix}~~~~~~~~~~~~~~~~~~~~~~~~~~~~~~~~~~~
\label{eq:eight}
\end{eqnarray}
We note that beam waist at focus appears now to be broadened by a factor $\sqrt{2}$ when comparing with the distribution of equation 13. Second we found that the confocal filtering by convolution with a Gaussian mode leads to a modified effective Jones matrix for the reflecting surface acting on the field as seen from the detector
\begin{eqnarray}
\bar{\bar{M}}_{Dx_0,y_0}=\begin{bmatrix}
r_p  & 0 \\
0 & r_s 
\end{bmatrix}+i\frac{x_0}{2l_f}\begin{bmatrix}
r'_p  & 0 \\
0 & r'_s 
\end{bmatrix}+i\frac{y_0}{2l_f}\frac{r_p+r_s}{\tan\theta_i}\begin{bmatrix}
0 & -1 \\
1 & 0 
\end{bmatrix}~~~~
\label{eq:nine}
\end{eqnarray}
With the confocal filtering, the Goos-Hänchen and the Imbert-Fedorov fields terms (i.e. the second and third terms of the r.h.s in the equation) are halved when compared to equation 12. The Jones matrix related to the polarizers and polarization leakage remain unchanged. With this last correction, we have now all the equations required in order to simulate the modal transformation induced by a reflecting surface acting on a polarized Gaussian beam and this for any arbitrary polarization and polarization leakage level. Finally the full scanning confocal mapping of the detected field is given by the analytical form
\begin{eqnarray}
\vec{{E}}_{D}(x_0,y_0)=\bar{\bar{A}}(\alpha) \bar{\bar{M}}_{Dx_0,y_0} \bar{\bar{P}}(\beta)\vec{{E}}_{D0,0}
\label{eq:nine}~~~~~~~~~~~~~~~~~~~~~~~~~~~~~~~
\end{eqnarray}
	
The first important result we are getting from equation (16) and the mirror matrix term in equation (15) is when the location of the fiber center and the focal spot axis coincide, namely for $x_0= 0$ and $y_0= 0$. In this case the result is the same as found in the simplifed plane wave analysis of section III and the equation of polarization cancellation given in equation 6 holds fortuitously. This is the case because the filtering function of the confocal arrangement eliminates the higher depolarizing modes. Without the confocal filtering, the normalized integrated total intensity in cross-polarization detected in wide field imaging of the focused point, or for collected with a wide core multimode fiber, is obtained using equations (10) and (13): 
\begin{eqnarray}
\frac{\iint \mid {\vec{\mathcal{E}}}_{f\perp}\mid^2 dx dy}{\iint \mid {\vec{\mathcal{E}}}_{0x,y}\mid^2 dx dy}=\frac{1}{4}\frac{\mid r_p+r_s\mid^2}{\tan^2\theta_i}\Big(\frac{\omega_f}{l_f}\Big)^2
\label{eq:nine}~~~~~~~~~~~~~~~~~~~~~~
\end{eqnarray}
For our experimental parameters and using equation (11), the collected depolarized field would limit the extinction to $1.56\times10^3$  and $2.1\times10^3$ for $p$- and $s$-polarized beam respectively. This demonstrates the key role of the confocal arrangement for the giant polarization extinction reached in the state of the art resonance fluorescence measurements \cite{vamivakas2009spin, kuhlmann2013dark,kuhlmann2013charge,kaldewey2018far}.\\
\indent The second significant result is illustrated, applying equation (16) on a purely $p$- or $s$-polarized beam measured in cross-polarization, by mapping the focused spot position $ (x_0, y_0)$ across the single mode fiber end. The result is an intensity map displaying two lobe maxima located at the fiber location at  $x_{0max}=0$ and $y_{0max}=\pm \omega_f$ above and below the plane of incidence. This is in complete agreement with our measurements as seen in Fig.~\ref{fig22} for silver. We confirmed quantitatively these findings for Bragg mirrors and thin-film based beam-splitter cubes. As seen in figure Fig.~\ref{fig33} and Fig.~\ref{fig44}, equation (16) maps closely the evolution of the mode transformation near cross-polarization condition for a metallic surface. In our experiment the material parameter of the high reflectivity surface that governs most of the effects we observed is the phase difference $\varphi_p-\varphi_s$. In particular for high reflectivity materials the intensity of the lobe maxima at cross-polarization are obtained from equation (16) at fiber location $(x_0,y_0)=(0,\pm \omega_f)$
\begin{eqnarray}
\mid {\vec{{E}}_{D}(0,\pm \omega_f)}/{\vec{{E}}_{D0,0}}\mid^2=\frac{1}{4e}\frac{\mid r_p+r_s\mid^2}{\tan^2\theta_i }\Big(\frac{\omega_f}{ l_f}\Big)^2~~~~~~~~~~~~~~~~~~~~~~~
\label{eq:nine}
\end{eqnarray}

In particular for high reflectivity materials, from equation 11 we have $\mid r_p+r_s\mid^2=4\cos^2 \Delta$. For a pure silver surface $\Delta=192.52^\circ$, the lobe intensity should be $5.9\times 10^{-5}$. For our independently measured values of $\Delta=102.48^\circ$ and $\Delta=100.7^\circ$,  we should be finding $2.35\times10^{-4}$ and $1.74\times10^{-4}$ for $p$- and $s$- polarization respectively. We measure typical lobe maxima in the range of 0.3 to 1.4$\times10^{-6}$. For a reason not yet elucidated, our maximum measured intensities are weaker than modeled. We believe that we are still missing a full quantitative understanding in the way the receiving fiber filters non Gaussian modes. Indeed the sensitivity of our setup should have permitted to detect the higher terms modes TEM$_{11}$ that have a symmetry $xy$. Such modes originate from the finite sized waist of the Gaussian beam making it naturally divergent \cite{PhysRevE.49.5778}. In fact, we can use the exact formalism developed above to show that such terms originate also from geometrical projections around the optical axis. This time the projection is not involving any reflecting surface but just the natural divergence of the beam before the lens, leading to a gradual phase-shift gained by each plane-wave component here again depolarizing naturally the beam. Applying the Fourier transform due to the focusing lens and keeping in mind the convolution imposed by the collecting fiber, we calculated that the expected clove shaped mode is  peaking at the four location $(x_0;y_0) = (\pm \omega_f;\pm \omega_f)$ with an intensity given by: 
\begin{eqnarray}
\mid \vec{{E}}_{D}(\pm \omega_f,\pm \omega_f)/\vec{{E}}_{D0,0}\mid ^2=\frac{1}{4e^2} \Big(\frac{\omega_f}{2~l_f}\Big)^4
~~~~~~~~~~~~~~~~~~~~~~~~~~
\label{eq:nine}
\end{eqnarray}
A result corroborated in ref \cite{PhysRevE.49.5778}. Using this expression for our experiment parameters, the mode peak intensity should be 3.8$\times10^{-7}$ a value that is well within our sensitivity range. It is a puzzling fact that we did not observe any trace of this TEM$_{11}$ signal. There is no doubt however that this mode is present as measured in  \cite{PhysRevE.49.5778} , this is why we believe that our understanding of the way the optical fiber is filtering the signal is not complete yet.

\begin{figure}
\includegraphics[scale=.4]{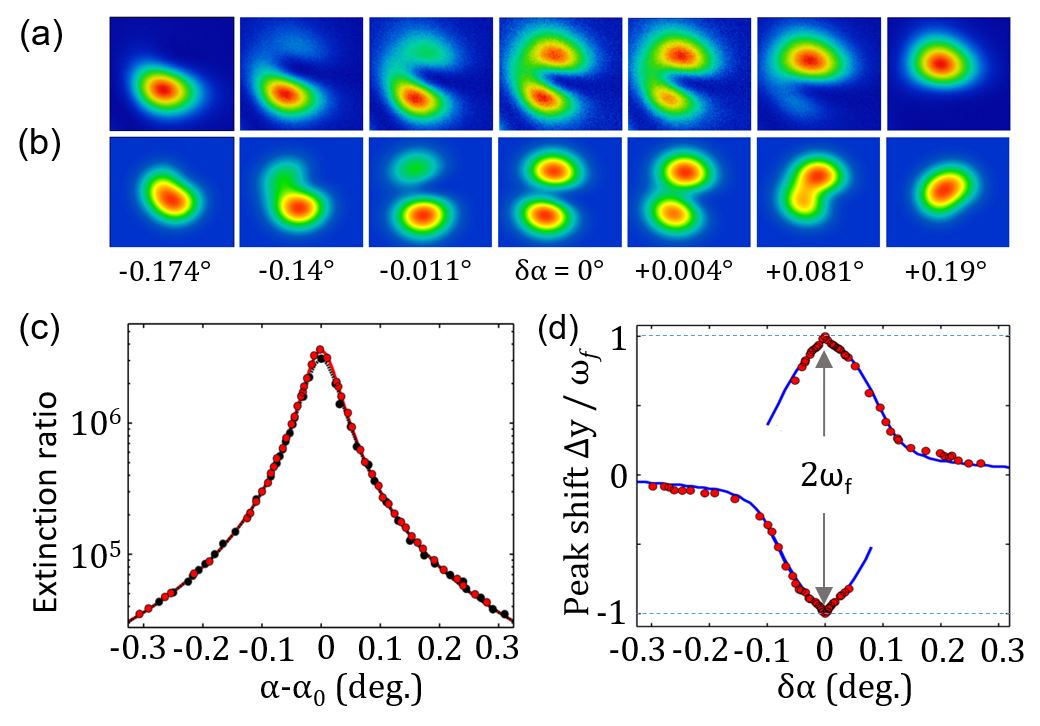}
\caption{\label{fig55} Measured (a) and simulated (b) evolution of the modal mapping through maximum extinction for reflectivity from the air side off a glass surface (BK7) in for $p$-polarization. (c) Red symbols: extinction ratio for different analyzer angles $\alpha$ shifts as explained in text. Black symbols: reference measurement with the analyzer placed just after the polarizers. The maxima beam peak-shift and splitting positions are shown and modeled in (d) in units of beam waist $\omega_f$.}
\end{figure}
Finally and for comparison we have also tested our model with a purely dielectric BK7 glass surface with reflectivity from the air-side near cross-polarization for the $p$-polarization. The results are shown in Fig.~\ref{fig55}. In this configuration as expected from the model discussed in the  section III indeed there is no shift $\alpha-\alpha_0$ between the condition of maximum cross-polarization for the dielectric and the reference measurement. As expected also from section III there is no effect of cancelation of the polarization leakage. The most remarkable difference is the way of the mode splitting evolves upon rotation of the analyzer. The absence of imaginary terms in $r_p$ and $r_s$ is the reason for this behavior. Here, we see not only the Imbert-Fedorov out of plane splitting at cross-polarization but also the appearance of the Goos-Hänchen mode showing a mixing that bends the beam shape along the plane of incidence. Fig.~\ref{fig55} shows in particular the evolution of the beam splitting near cross-polarization condition which is completely different from what is seen for metals such in in Fig.~\ref{fig33}d and Fig.~\ref{fig44}d. Because $Im(r_s r_p^*)= 0$  for the reflectivity from the air side of a dielectric there is no term linear in $\delta\alpha$ near cross-polarization conditions as also seen from the experiments. For a dielectric an analytical solution for the location of the lobes intensity maxima for very small analyzer rotation angle $\delta\alpha<<1$ near cross-polarization condition shows the quadratic evolution observed in our measurement. 
\section{Conclusions}
\label{Conclusions}
 In conclusion, we have exposed a systematic experimental method based on a confocal microscopy arrangement to obtain a giant enhancement in dark-field cross-polarization extinction and this by up to 3 orders of magnitudes and possibly beyond. We found that the effect exploits the material properties of a surface or interface under condition of total internal reflectivity in particular. In more general terms, the effect was found to be fully governed by the phase difference, such as induced by beam-splitter cubes and Bragg mirrors, between the reflectivity of light components polarized in- and out-of-plane of incidence. Modeling this effect led us to simulate and map in minute details the transformation of Gaussian beams near cross-polarization into Imbert-Fedorov higher modes, a physics governed essentially by the finite divergence of a Gaussian beam reflecting off a surface. This work opens the way to methodical design of sensitive laser resonant fluorescence microscopes with extreme background extinction, for a broad range of applications in quantum optics and solid-state physics. The 
new methods developed for this work can also be applied for measuring material optical properties.\\
\begin{acknowledgments}
We thank R.~Warburton, M.~Kroner, L.~Novotny, and C.~Schaefermeier for useful discussions.
Part of this project has received funding from the European Union’s Horizon 2020 research and innovation programme under the Marie Skłodowska-Curie grant agreement No 721394 ITN 4PHOTON.
\end{acknowledgments}


\end{document}